\documentclass[twocolumn,prl,showpacs,preprintnumbers,amsmath,amssymb]{revtex4}

\usepackage{epsf}
\usepackage{epsfig}
\usepackage{graphicx}
\usepackage{dcolumn}
\usepackage{bm}

\newcommand{\J}{{\rm\bf J}} 
\newcommand{\Jmag}{{\rm J}} 

\begin{document}


\frenchspacing

\title{Easily repairable networks}

\author{Robert S.\ Farr$^{1,2}$, John L. Harer$^{3}$ and Thomas M.\ A.\ Fink$^{1,4}$ }
\affiliation{$^{1}$London Institute for Mathematical Sciences, 35a South St, London W1K 2XF, UK}
\affiliation{$^{2}$Unilever R\&D, Colworth Science Park, MK44 1LQ, Bedford, UK}
\affiliation{$^{3}$Department of Mathematics, Duke University, Rayleigh-Durham, NC, USA}
\affiliation{$^4$Centre National de la Recherche Scientifique, Paris 75248, France}
\email{robert.farr@unilever.com, harer@math.duke.edu, tf@lims.ac.uk}

\date{\today}

\begin{abstract}
\noindent
We introduce a simple class of distribution networks 
which withstand damage by being repairable instead of redundant.
We prove a lower bound for the expected cost of repair, 
and show that for networks on the square and triangular lattice, this bound is achievable
and results in a network with exactly three levels of structural hierarchy. 
We extend our results to networks subject to repeated attacks, in which the repairs themselves must be repairable.
We find that, in exchange for a modest increase in repair cost, such networks are able to withstand any number of attacks.
\end{abstract}

\pacs{84.40.Ua, 84.30.Jc, 89.40.Cc}

\maketitle

\noindent 
Increasing the resilience of infrastructure networks to natural and man-made disasters is a topic of the highest political concern \cite{unisdr}.
In recent years we have witnessed the devastating impact of both natural 
calamities (tsunamis, hurricanes and earthquakes) 
and man-made threats (sabotage, accidents).
The human loss is exacerbated by the collapse of
distribution networks (power and water), communication grids (cellular and {\sc www}) and transport networks (road and rail) \cite{Tero}, even though 
the fraction of each network that is damaged may be small.

To safeguard against the threat of disasters, many researchers and 
funding agencies have focused on robustness, 
whereby damage is absorbed due to internal redundancy.
Robustness tends to be the strategy adopted by biological networks, 
such as the circulatory and nervous systems and leaf venation \cite{Katifori},
which also must function reliably under environmental insults \cite{Benfey}.
Robustness, however, is not the only strategy for increasing resilience. 
In a recent announcement, the EU science 
agency appealed for ``resilience concepts [that] take into account the 
necessity to anticipate, plan and implement a substitution process in a 
crisis or disaster, aiming to deal with a lack of\ldots capacities 
necessary to assume the continuity of basic functions and services,
until recovery from negative effects and return to the normal 
situation" \cite{EC}.
This ``substitution process," or workaround, involves finding a short-term fix until the damaged part itself can be repaired. 
When the typical cost of a workaround (averaged over all possible failure modes) is low, we say that the system is repairable.
Repairability in this sense is the subject of this Letter.

For concreteness, we define the two resilience strategies thus: 
A network is \emph{robust} if, after an error in part of it, it is able (or more likely) to function normally on account of internal redundancy.\ 
A network is \emph{repairable} if, after an error in part of it, it is able (or more likely) to function normally on account of intervention in other parts of it.

Before we consider network resilience, we briefly outline optimal 
infrastructure networks.
The simplest models take connectivity to be the sole determinant of function.
Such models are appropriate for certain networks under light load, 
such as roads, electricity supply \cite{Boruvka} and communication 
networks \cite{History}.
Here the network cost typically grows with the total length of the edges, 
and optimal solutions in a constant environment are minimal spanning trees \cite{Boruvka}.
As a network becomes more heavily used, connectivity alone is no longer 
sufficient, and the capacity of the edges must also be considered. 
For models derived from resistor networks, efficiency translates to 
minimum power dissipation.  If one associates a cost $R^{-\gamma}$ with 
each resistance $R$ and specifies a total cost, then for planar networks 
with $\gamma<1$, loopless geometries are known to be 
optimal \cite{Bohn,Durand}. More generally, this approach provides an 
explanation for fractal branching networks in biology, and ultimately 
for allometric growth laws \cite{West}.

So much for optimal networks in a static environment. In the presence of 
unexpected events, the traditional approach to maintaining function is the 
introduction of redundancy (localized, such as extra paths from source to 
sink, or distributed, such as checksums in digital data). 
For models of the internet, where simple connectivity suffices, the 
exponent of the distribution of node degrees \cite{Barabasi} 
is the key parameter, both for random \cite{Cohen2000} 
and directed \cite{Cohen2001} attacks. For planar resistor networks 
with $\gamma<1$ suffering local damage or under fluctuating loads, 
the designs that emerge from numerical optimization have a hierarchical 
loop structure over many length scales \cite{Katifori}.

\begin{figure}[b!]
\begin{center}
\includegraphics[width=1\columnwidth]{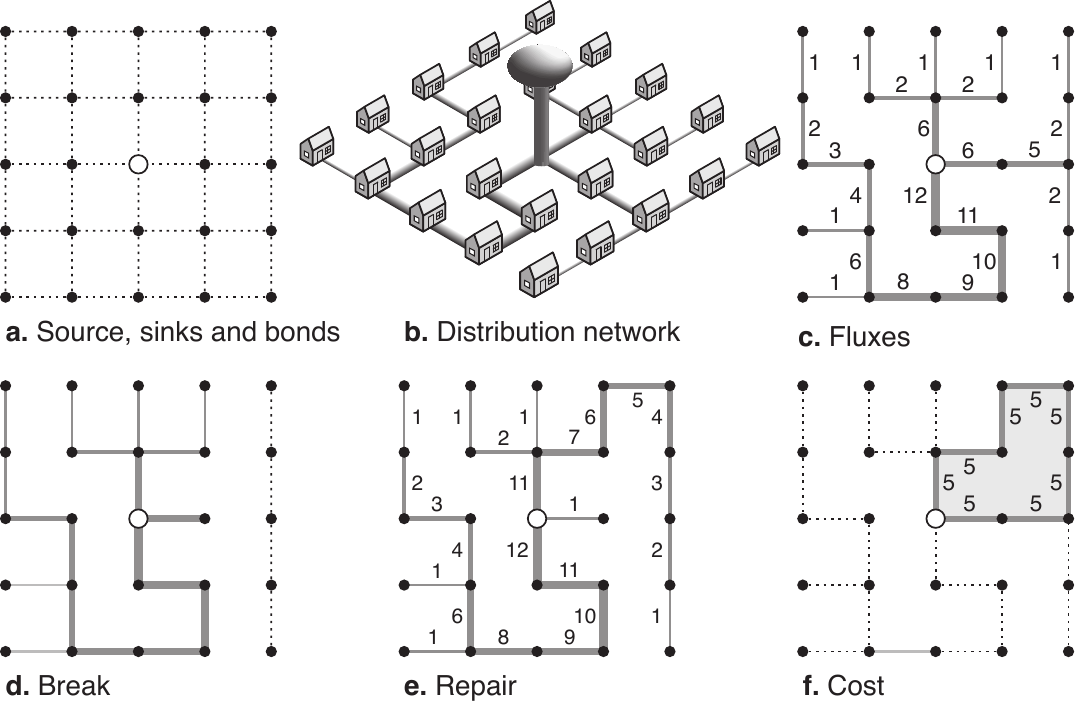}
\caption{
\label{damagerepair}
A distribution network, and the process of damage and repair.
 {\bf b}. Each house requires one unit of water per day from a tower.
The flux into each house is equal to the number of houses downstream 
from it, plus one. Each day a random pipe breaks, and a new pipe must be 
built elsewhere to reconnect the cut off houses.
Moreover, some of the pipes upstream and downstream from the new pipe must be resized 
to accommodate the change in demand.
What network minimizes the expected total change in flux, and thus pipe altered, per break?
 {\bf a}. The source, nodes and bonds (possible edges).
 {\bf b}. The tree of edges connecting the nodes to the source.
 {\bf c}. The fluxes $\J_{i}$ at each edge.
 {\bf d}. An edge is broken, splitting the tree in two.
 {\bf e}. A new bond is made an edge, reconnecting the subtrees, and the 
updated fluxes $\J'_i$.
 {\bf f}. The changes in fluxes $|\J'_{i}-\J_{i}|$.
They vanish everywhere except on the loop through the broken and repairing edges.
The cost of repair is the flux through the broken edge times the remaining loop 
length: $c$ = 5 $\cdot$ (8-1) = 35.
}
\end{center}
\end{figure}

The above work has focused on network robustness, whereas we 
want to understand network repairability. 
In doing so, we wish to capture the constraints on real infrastructure networks, 
in particular the cost of capacity rather than just connectivity.
Resistor networks account for this in a natural way, but have the 
disadvantage that analytic results are hard to obtain. 
We therefore propose a model of intermediate complexity, 
where the required capacity of each edge is proportional to the number 
of downstream nodes it has to serve. An example of such a model is shown 
in Figure \ref{damagerepair}b, where houses are supplied with 
water from a central tower.  
This model captures essential 
features of networks under non-trivial loads, but is sufficiently 
tractable to solve interesting cases exactly.

In this Letter, we do the following three things, each of which corresponds to a separate section: 

\hangindent=1.3em
\noindent 1.\ We introduce a model of repairable networks in which a break at 
edge $i$ can be mitigated by adding an edge at $j$, and show that the cost 
of repair is the flux at $i$ times the length, less 1, of the loop 
through $i$ and $j$.

\hangindent=1.3em
\noindent 2.\ We introduce the concept of an easily repairable 
network ({\sc ern}), which minimizes the expected cost of 
repair $\langle c \rangle$ after a single attack.
We prove that {\sc ern}s have exactly three levels of structural hierarchy. 

\hangindent=1.3em
\noindent 3.\ When attacks are sufficiently numerous to strike the same 
place repeatedly, the repairs themselves must be easily repairable.
To address this, we describe \emph{steady-state} {\sc ern}s, able to 
withstand any number of attacks in exchange for a modest increase 
in $\langle c \rangle$.

\noindent{\it Model of repairable networks.---}
For some systems, mistakes are reversible: the unintended change can be reverted
or the broken part can be replaced like-for-like (an on-screen typo; a dropped pen; a flat tire).
More often than not, however, the error is irreversible or the repair time is unacceptably long 
(a printed typo; a missing ingredient; a jet engine failure).
In these cases, it is necessary to find a workaraound; 
that is, to restore the broken part by intervening in other parts. 

As a model of the latter, suppose we need to continuously transport a commodity or information
from a single source to a collection of $N$ nodes, each of which consumes 
the substance at a unit rate. We imagine that the source and $N$ nodes 
form the vertices of an underlying lattice $L$, which we take to be either
square ($L_\square$, illustrated in Figure \ref{damagerepair}), or triangular with 
a hexagonal boundary ($L_\triangle$). We assume the diameter is $l$ nodes, where $l$ is odd. 
The total number of vertices is $l^{2}$ ($L_\square$) 
or $\frac{3}{4}(l^2-1) + 1$ ($L_\triangle$), so that $N_\square=l^{2}-1$ 
and $N_\triangle = \frac{3}{4}(l^2-1)$. The bonds of $L$ are the 
possible conduits for transport, but not all of these bonds 
will be used in practice. The ones that are we call {\it edges}, and 
the ones that are not {\it passive edges}.  Each edge $i$ carries a 
signed flux ${\J}_{i}$, such that the flux leaving the source is $N$, 
and the net flux into each of the $N$ nodes is 1.  We assume (without 
loss of generality, as we shall see later) that the network of edges is 
a tree $T$ (Figure \ref{damagerepair}b). This loop-free property allows us 
to assign fluxes on $T$ unambiguously (Figure \ref{damagerepair}c).
It also means there must be $N$ edges, because into each node flows 
exactly one edge.

We now consider what happens when the network is broken and then repaired.
In our model, a break consists of disabling a single edge $i$ ({\it i.e.}, 
edge $i$ becomes passive; passive edges have zero flux), which disconnects 
the tree $T$ into two disjoint trees (Figure \ref{damagerepair}d).  
We then proceed to repair the network by adding an 
edge $j\neq i$ such that the new network is once more a tree, with a new set of 
fluxes $\{ {\J'}_{i}\}$ (Figure 2e). 
We define the cost (intervention required) $c_{ij}$ from this break-repair operation 
as the sum of the absolute changes in flux, but omitting the flux in the broken 
edge (we do not pay for the attack):
\begin{equation}\label{cij}
c_{ij}=-|{\J}_{i}|+\sum_{k\, {\rm an \, edge}}|{\J'}_{k}-{\J}_{k}|,
\end{equation}
This makes sense: if we imagine a fluid to flow along small unit flux pipes 
in parallel, or cars to travel along unit flux lanes of a highway, $c$ 
is the number of pipes or lanes to add or remove (where reversing involves 
adding and removing). 
Note in particular that we pay if capacity is reduced; 
this is a valid strategy if there is an ongoing maintenance cost attached to the capacity of each edge (\emph{e.g.}, metabolism in biological tissues), 
so that it is rational to pay up-front to eliminate this. 
This is the natural choice from a mathematical perspective, but may not always be physically realistic.
We thus consider the alternative model where spare capacity is free; here the costs turn out to be proportional our original model solutions.

\begin{figure}[b!]
\includegraphics[width=1\columnwidth]{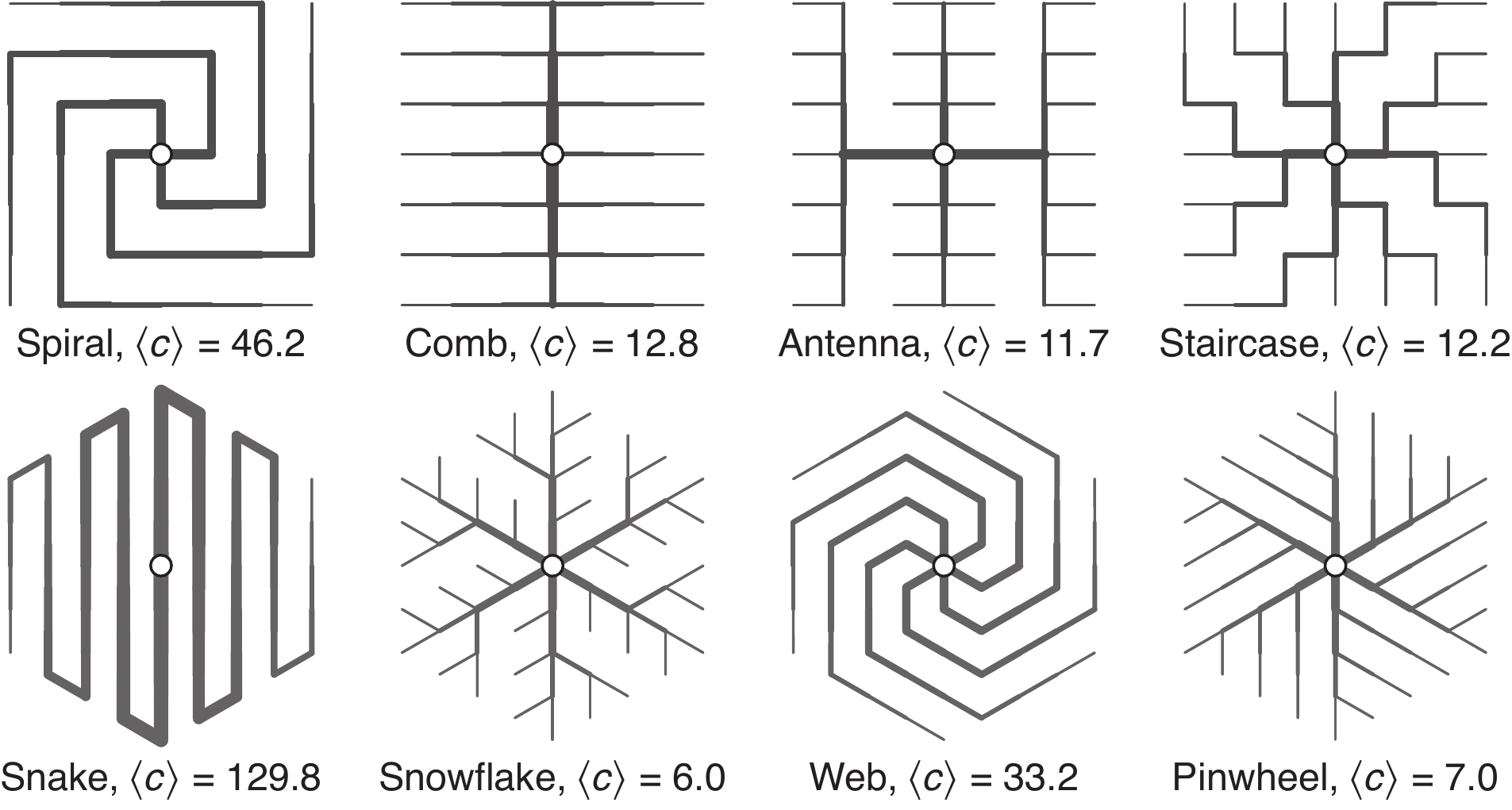}
\caption{\label{strategies}
Different strategies for building a distribution network and their expected cost of 
repair $\langle c \rangle$. Top: square lattice. Bottom: triangular lattice.
}
\end{figure}

In order to evaluate $c_{ij}$, suppose we have any two valid networks
$T'$ and $T''$ with the same source location.
If we subtract the fluxes in $T''$ from those in $T'$, the resultant pattern of 
fluxes must have no sources or sinks.
Therefore it must either vanish, or be a sum of closed flux loops.
Now consider our original network $T$.
If we make a new network containing the original edges of $T$ and the added
edge $k$, this network is no longer a tree, but contains exactly one loop, of 
length $d_{ij}$.  When we take the difference of the fluxes in the original and 
repaired networks, this is the only path which can have a non-zero flux 
(Figure \ref{damagerepair}f), which must be the original flux 
at $i$ (since after repair, the flux in this edge is zero). 
Consequently, the cost of repair is equal to the length of the loop 
through $i$ and $j$ less $1$ (we omit the broken edge), times the flux at $i$:
\begin{equation}
	c_{ij} = \Jmag_{i} (d_{ij}-1),
	\label{loop}
\end{equation}
where $\Jmag_i = \vert \J_i \vert$.\
Now a broken network can be mended in a number of ways, corresponding to 
different choices of the bond $j$ used to repair it.
We desire the {\it cheapest} repair, which is the one with the least cost: 
\begin{equation}
	c_i = {\Jmag}_{i} \min_{j} (d_{ij}-1).\
	\label{Delta_min}
\end{equation}
Hereafter we take `repair' to mean `cheapest repair', and call the loop 
that arises from considering both the broken edge and the repair a {\it dormant loop}.
Because the cost of repair $c_i$ depends on the particular edge $i$ that is 
broken, which is unknown, we want the {\it expected cost of repair}:
\begin{equation}
	 \langle c \rangle = {1\over N} \sum_{i \, {\rm an \, edge}} {\Jmag}_{i}  \min_{j} 
   (d_{ij}-1).
	\label{Cdef}
\end{equation}
What network minimizes $\langle c \rangle$? 
Figure \ref{strategies} shows some examples of networks, with associated values 
for $\langle c \rangle$.

\begin{figure}[b!]
\includegraphics[width=0.652\columnwidth]{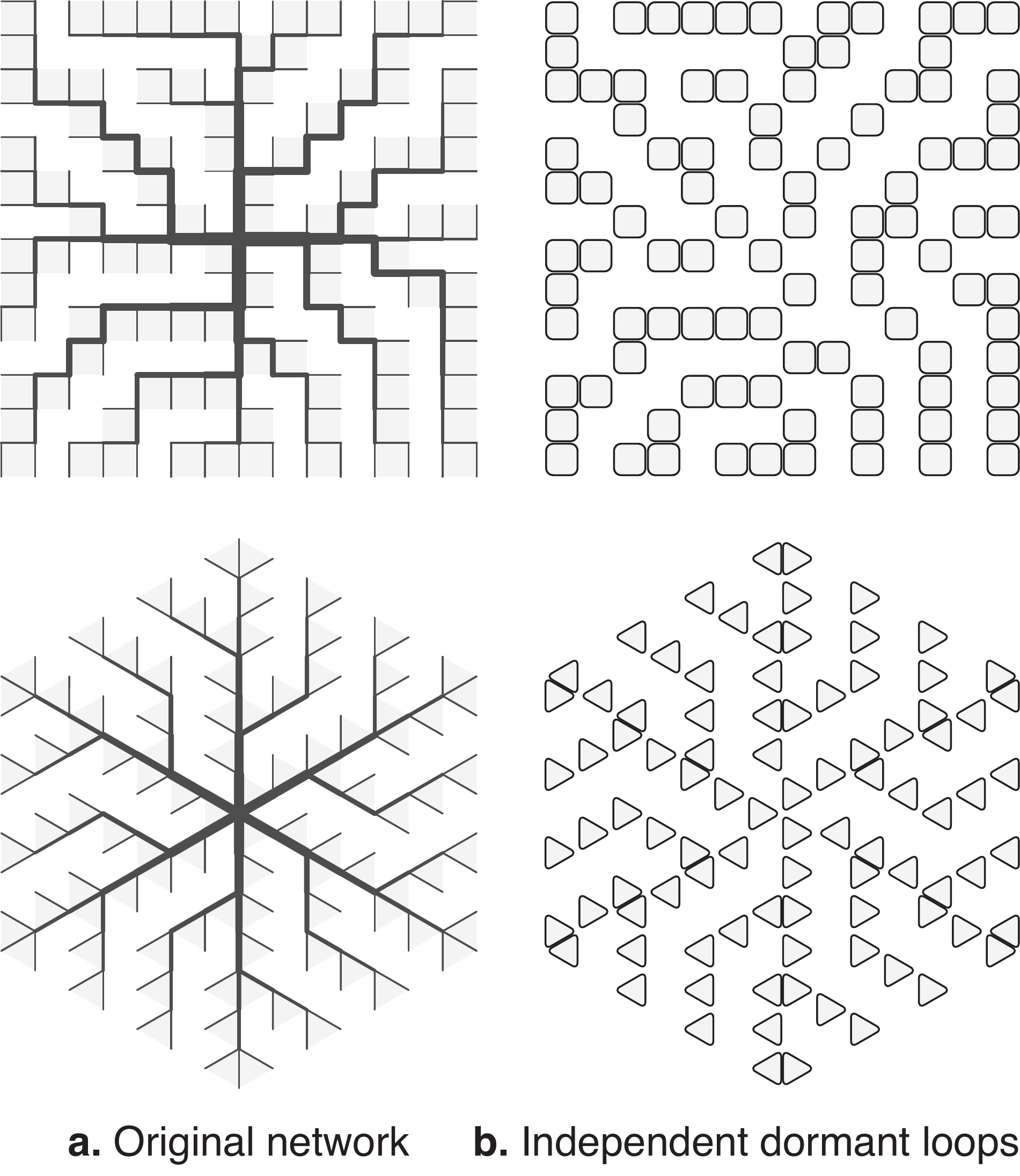}
\caption{\label{optimal_one}
Networks that are easy to repair after one break.
The original networks (left) and the configuration of overlapping dormant loops (right).
}
\end{figure}

\noindent{\it Easily repairable after one break.---}
The expected cost of repair $\langle c \rangle$ will be a minimum if 
two conditions are satisfied: the mean absolute flux 
$\langle\Jmag\rangle \equiv N^{-1}\sum J_i$ in the original network is
a minimum; and all the individual shortest loop lengths $d_{ij}$ are minimal.
Without knowing to what extent these conditions are independent, we first
ask: what networks minimize $\langle\Jmag\rangle$?
To answer this, we note that every node must be connected to the source 
by some path, and along this path flows a unit of flux. There may also be 
confluent fluxes to other nodes flowing in the same edges, but we can 
conceptually treat this flux separately, even if in practice we do not keep track of all the individual streams.
Accordingly the mean flux $\langle \Jmag \rangle$ is $1/N$ times the 
sum of the lengths of all these streams. Therefore a minimum 
of $\langle \Jmag \rangle$ is achieved when each of these paths is a 
geodesic (shortest path) of the graph $L$ between the node 
and the source. Such geodesics in general will not be unique, but
at least one network composed of these geodesics 
must be a tree, because any loop can be broken by removing an
edge at its greatest distance from the source and diverting the 
incoming streams into one side of this loop.  
If the source is at the center of the lattice, we find
$\langle\Jmag\rangle_{\min}=l/2$ 
for a square lattice ($L_\square$) and $l/3$ for a triangular lattice ($L_\triangle$).
Since the minimum loop length $d_{ij}$ on $L_\square$ and $L_\triangle$
is $4$ and $3$, we find
\begin{equation} 
\langle c\rangle_\square \geq 3l/2 \quad {\rm and}  \quad \langle c\rangle_\triangle \geq 2l/3.
\end{equation} 
Because both of these criteria can be met simultaneously, as Figure \ref{optimal_one} exemplifies, 
the above bounds can be achieved. We call the solutions, which are not unique, easily 
repairable networks ({\sc ern}s). 

We now consider the case where we do not have to pay to immediately reduce capacity.
For an {\sc ern} on $L_\triangle$, the expected cost of repair is identical in both cases, because none of the fluxes change direction, and we do not pay for the break itself. 

The structure of {\sc ern}s on regular lattices is remarkable
in that they have exactly three levels of hierarchy: 
connected to the source are primary arms (1-arms), 
from which branch secondary arms (2-arms),
from which branch terminal hairs of length $1$ (3-arms). 
In this sense {\sc ern}s loosely 
resemble crinoids: marine animals with cilia-covered arms connected 
to a primary stalk. The architecture of {\sc ern}s is in contrast to 
robust resistor networks, which have a hierarchical branching structure 
over many generations \cite{Katifori}.
The steps to our proof of this limited hierarchy are as follows.
(i) A 1-arm must lie along a coordinate axis; were it otherwise, the path to at least one of the nodes on that axis would include a bend and not be a geodesic. 
(ii) When a 2-arm splits, there can be only two daughter branches (including the 2-arm); 
there are only two directions that are geodesic, namely, the two away from the origin. 
(iii) In any 2-arm split, the two daughter nodes cannot both split; if they did, a closed loop would be formed.
(iv) The middle node of two consecutive edges must have a split; were it otherwise, the distal edge could not be part of a dormant loop of minimal size.

\noindent{\it Easily repairable after many breaks.---}
So far we have described networks that are easily repairable after a 
single break. What happens when there is a series of breaks?
As we indicated in the introduction, 
we suppose that a break cannot be repaired immediately but is open to intervention later on 
(for example, if there there is another attack in the neighborhood).
Clearly, repairing the first break (at an expected cost of $3l/2$ or $2l/3$) leaves the network fully functional.
However, there is a more subtle effect: the repairs themselves may not be optimally repairable.
Consequently, as the number of breaks increases, the network degrades, 
and the mean cost of repair $\langle c\rangle$ goes up.

\begin{figure}[b!]
\includegraphics[width=1.0\columnwidth]{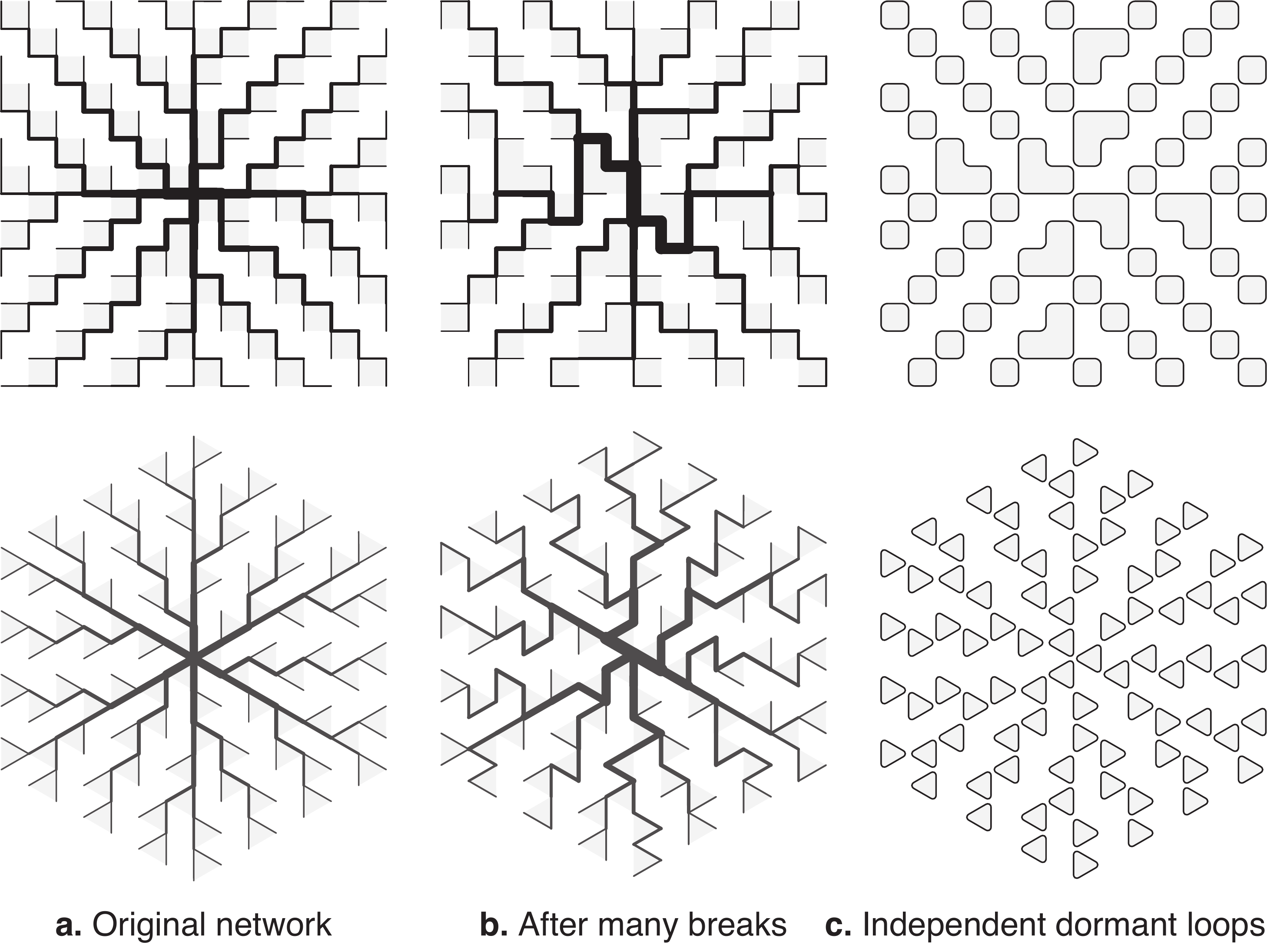}
\caption{\label{optimal_many}
Networks that are easy to repair after many breaks.
The original networks (left), the networks after many break and repair cycles (center), 
and the configuration of independent dormant loops (right).
}
\end{figure}

\begin{figure}[t!]
\includegraphics[width=0.85\columnwidth]{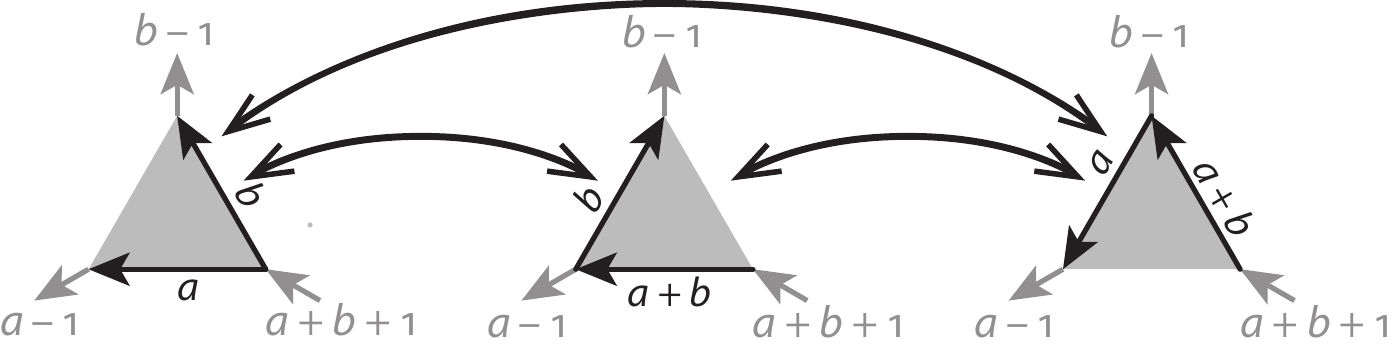}
\caption{ \label{spins}
The three orientations of a dormant loop. The left orientation is geodesic (the 
flux follows the shortest path to each node), whereas the other two are not.
Each break and repair sends the dormant loop from one orientation to another.
The total flux along the geodesic dormant loop is $a+b$. After many 
break-repair cycles, when the orientations are equally likely, the 
mean total flux is $\frac{4}{3}(a+b)$.
Thus at steady state, the mean flux $\langle J(\infty) \rangle$
is 4/3 of that before any breaks, $\langle J(0) \rangle$.
} 
\end{figure}

How do we rectify this?
The structure of an {\sc ern} is such that 
(i) it is geodesic, thereby minimizing $\langle J\rangle$; and 
(ii) each of its edges is part of a minimal dormant loop, thereby minimizing $d$.
Network degradation is due to two effects: the failure of 
the repaired network to be geodesic; and the increase in dormant loop length.
The former is because the edges in a minimal latent loop can have
only 4 ($L_\square$) or 3 ($L_\triangle$) orientations, and not all of these are 
geodesic; since repair is performed on the only passive edge, eventually 
the orientations will occur with equal frequency. The latter is the result 
of dormant loops that share an edge; and when an edge belonging to two 
dormant loops is broken, repairing it coalesces the two loops into one larger 
dormant loop. Therefore to optimize networks for cheap repair under 
multiple breaks, we must ensure that the dormant loops are independent, 
having no edges in common.  Figure \ref{optimal_many} shows that for a 
triangular lattice, this can be achieved;
for the square lattice, there will always be at least a small fraction of
dormant loops sharing an edge, and in fact these dominate the change 
in $\langle c\rangle$ after many breaks. Figure \ref{spins} shows the 
calculation for the average repair cost per break after many breaks.

The final result in the many-break limit is
\begin{equation}
\langle c\rangle_\square \rightarrow 7l/4+O(1) \quad {\rm and} \quad \langle c\rangle_\triangle \rightarrow 8l/9.
\end{equation}
Thus in both cases, the networks are able to withstand any number of 
breaks in exchange for a modest increase in the expected repair cost.
We call the solutions {\it steady-state} {\sc ern}s.
Their design aims to achieve three properties (all being simultaneously possible
for the triangular lattice):
(i) the initial network is geodesic;
(ii) the dormant loops are minimal;
(iii) the dormant loops do not overlap.

We briefly consider the case where we do not have to pay to immediately 
reduce capacity.
Analysis for $L_\triangle$ similar to that in Figure \ref{spins} (but
now involving 6 possible states) reveals 
that steady-state {\sc erns} have an asymptotic expected cost 
$\langle c\rangle_{\triangle}\rightarrow 5l/9$.

\noindent{\it Discussion.---}
There are two aspects of repairability we have not addressed so far.
If the edges are not all of equal length, we can generalize our model by 
letting $d_{ij}$ be the physical length of the loop, 
not just the number of edges it contains. 
Then our model would encompass networks whose underlying lattice is irregular. 
For small perturbations from the regular lattice $L$ there will be no new 
solutions (since moving away from a solution that is optimal on $L$ incurs a cost of order unity); 
however, the perturbation is likely to break the degeneracy between 
different optimal solutions on the original $L$. 
We also conjecture that it is possible to optimize networks which are 
repairable to damage across multiple length scales, 
which may involve the presence of minimal dormant loops of different sizes, 
{\it e.g.}, on the network coarse-grained by $2\times 2$, $4\times 4$ and so on.

We believe our model of distribution networks captures essential features 
of their real-world analogs, in a form which is simple enough to be 
analytically tractable.  
It suggests that the structure of {\sc ern}s and 
steady-state {\sc ern}s embodies useful design directions for engineering 
applications, such as resource distribution, smart electricity 
grids \cite{Amin} and communication networks. 
More generally, it helps quantify the concept of repairability, and offers a framework for extending our understanding of repairability as an alternative to robustness in achieving resilience.

\noindent{\it Acknowledgements.---}
The authors acknowledge support from 
the Defense Threat Reduction Agency,
the Boston Consulting Group
and EU FP7 ({\sc growthcom}).

\end{document}